\documentclass[letterpaper, 10 pt, conference]{ieeeconf}
\usepackage{filecontents}
\usepackage{cite}
\usepackage{algorithmic}
\usepackage[ruled,vlined]{algorithm2e}
\usepackage{amsmath}
\usepackage{algorithmic}
\usepackage{algorithm2e}
\usepackage{float}
\usepackage{subfloat}
\usepackage{graphicx}
\usepackage{subcaption}
\begin{document}
\title{\textbf{Hyperspectral Unmixing by Nuclear Norm Difference Maximization based Dictionary Pruning}}
\author{Samiran Das $^{1}$, Aurobinda Routray $^2$ and Alok Kanti Deb $^{2}$\\% <-this % stops a space
{1} Advanced Technology and Development Center, Indian Institute of Technology Kharagpur\\
{2} Department of Electrical Engineering, Indian Institute of Technology Kharagpur}
 \thanks{*This work was not supported by any organization}
\maketitle

\begin{abstract}
Dictionary pruning methods perform unmixing by identifying a smaller subset of active spectral library elements that can represent the image efficiently as a linear combination. This paper presents a new nuclear norm difference based approach for dictionary pruning utilizing the low rank property of hyperspectral data. The proposed workflow calculates the nuclear norm of abundance of the original data assuming the whole spectral library as endmembers. In the next step, the algorithm calculates nuclear norm of abundance after appending a spectral library element with the data. The spectral library elements having the maximum difference in the nuclear norm of the obtained abundance matrices are suitable candidates for being image endmember. The proposed workflow is verified with a large number of synthetic data generated by varying condition as well as some real images.
\end{abstract}

%\begin{IEEEkeywords}
%Hyperspectral Unmixing, Dictionary Pruning, Nuclear Norm Maximization, Sparse Unmxing
%\end{IEEEkeywords}

% \fi
\IEEEpeerreviewmaketitle
\section{Introduction}
Unmixing process is pivotal to accurate object identification and land cover mapping \cite{richards1999remote} from remotely sensed hyperspectral images. Hyperspectral images capture  reflectance pattern of a large number of spectral bands in the electromagnetic range from visible to infra-red region. The rich spectral information enables very accurate classification and object identification. Spectral unmixing process \cite{chang2013hyperspectral}-\cite{keshava2003survey} is used to identify the underlying spectral reflectance pattern of the objects or the endmembers present in the image and calculate abundance of the endmembers. Unmixing process can be unsupervised or supervised. Unsupervised unmixing process is a difficult and constrained blind source separation problem \cite{shippert2003introduction}, where signal sources (endmember matrix) and mixing coefficients or abundance of endmembers are estimated. However, spectral library based methods calculate the sparse abundance of matrix considering the whole spectral library as the endmembers or identify a subset of spectral library elements present in the image. 

Overall the paper is organized in the following sections- Section II gives an overview of spectral library based unmixing and introduces the existing algorithms, Section III describes the proposed nuclear norm minimization based dictionary pruning, Section IV includes the results obtained on simulated and real images, whereas, Section V discusses conclusion and future scope.
\section{Spectral Library based Unmixing}
According to linear mixing model the reflectance of the $i-th $ pixel can be written as-\\
$x_{i}=m_{i}D+w_{i}$\\
The hyperspectral image can be written as-
\begin{eqnarray}
X=MD+W
\end{eqnarray}
where, $X=\left[ x_{1}, x_{2}, .,., x_{N}\right ]$ is the hyperspectral image containing total $N$ pixels\\
where, $D=\left[ d_{1}, x_{2}, .,., d_{K}\right ]$ is the spectral library containing reflectance pattern of total $ K $ elements\\ 
 $M\epsilon {R^{N\times K}}$ is the mixing matrix or abundance matrix\\ 
 and $W\epsilon {R^{N\times L}}$ is the noise and residual term\\
Dictionary pruning algorithms identify the pruned dictionary $\hat{D}$ which contains a relatively lower number of elements. The resulting abundance matrix of the pruned library $\hat{D}$ has lower level of sparsity as compared to the abundance matrix of the original spectral library. 
\begin{eqnarray}
X=\hat{M}\hat{D}+\hat{W}
\end{eqnarray}
where, $X=\left[ x_{1}, x_{2}, .,., x_{N}\right ]$ is the hyperspectral image containing total $N$ pixels\\
where, $\hat{D}=\left[ \hat{d_{1}}, \hat{d_{2}}, .,., \hat{d_{R}}\right ]$ is the spectral library containing reflectance pattern of total $ R $ elements\\
$\hat{M}=\left[ \hat{m_{1}}, \hat{m_{2}}, .,., \hat{m_{R}}\right ]$ is the mixing matrix or abundance matrix of the pruned library\\ 
$\hat{M}\epsilon {R^{N\times R}}$
 and 
 $\hat{W}=\left[ \hat{w_{1}}, \hat{w_{2}}, .,., \hat{w_{N}}\right ]$ is the noise and residue term\\$W\epsilon {R^{N\times L}}$ is the noise and residue term.

In traditional unsupervised constrained unmixing, both estimation of the number of endmembers and estimation of signal sources are challenging in the presence of noise and it requires a huge amount of computation. Dictionary pruning algorithms overcome some of these difficulties posed by unsupervised unmixing.

Existing dictionary pruning schemes for unmixing generally include- MUSIC-collaborative sparse regression (MUSIC-CSR)\cite{iordache2014music}, robust MUSIC-dictionary aided sparse regression (RMUSIC-DANSER)\cite{fu2016semiblind}, subspace Matching pursuit (SMP)\cite{shi2014subspace}, compressive sampling matching Pursuit (CoSaMP), simultaneous orthogonal matching pursuit (SOMP), futuristic greedy algorithm \cite{akhtar2015futuristic} etc.
MUSIC-CSR algorithm\cite{iordache2014music} performs dictionary pruning by multiple signal identification and classification technique, whereas, the inversion process is performed by sparse regression. Robust MUSIC-dictionary adjusted non-convex sparsity encouraging regression (RMUSIC-DANSER)\cite{karhunen1991robust} proposes an improved noise robust version of this algorithm, which accounts for variability of spectral and mismatch between spectral library elements with actual image endmembers. 

Some of the existing dictionary pruning based greedy algorithms include- Orthogonal matching pursuit (OMP) \cite{tang2014sparse}, OMP star \cite{akhtar2015futuristic}, simultaneous orthogonal matching pursuit (SOMP) \cite{tropp2006algorithms}, subspace matching pursuit (SMP) \cite{shi2014subspace}, compressive sampling matching pursuit (CoSaMP) are greedy algorithms for dictionary pruning. These algorithms find the best matching projections of multidimensional data onto an over complete dictionary. These greedy pursuit algorithms differ in coefficient selection, residue update procedure and projection of the data on each individual atoms of the dictionary. Subspace matching pursuit (SMP) algorithm\cite{shi2014subspace} assumes that a particular endmember is present in multiple pixels and selects the particular spectral library elements iteratively. Orthogonal matching pursuit algorithm (OMP) \cite{akhtar2015futuristic} finds the spectral library element that has the most similarity with the residual. This element from the spectral library is stored and the residual is updated. This process is continued until the stopping criteria is met. Mostly used convergence criteria are- Residual goes below a threshold value or the rate of change of the norm of the residual matrix at successive iterations is very low or the number of iterations is below the maximum number of iterations specified by the user. 
% * <itzsamirandas@gmail.com> 2017-08-06T16:29:13.814Z:
%
% ^.
However, these dictionary pruning schemes schemes have certain limitations-
\begin{itemize}
\item{Most of the algorithms tend to overestimate the pruned dictionary.} 
\item{Some of these algorithms require significant amount of computation.}
\end{itemize}

High mutual coherence of spectral library poses major difficulties in spectral library based unmixing. Mutual coherence of spectral library is the maximum cosine distance between any two spectral library elements. Mutual coherence of spectral library is defined as-
\begin{equation}
\mu \left ( D \right )= \arg\max_{{1\leq{k}},{m\leq{j}},{k\neq{m}}}\frac{\left | {d_i{}^{T}}d_{j}\right |}{{{\left | {d_{i}} \right |}_2}{\left | {d_{j}} \right |}_2}
\end{equation}
%\IEEEpubidadjcol
\section{Dictionary Pruning by Nuclear Norm Difference Maximization}
Any hyperspectral image is a latent mixture of very few spectral library elements. As a consequence, the abundance matrix obtained by considering the whole spectral library as endmember, is highly sparse, because, it contains a large number of zero columns. The sparse inversion process minimizes the reconstruction error while maximizing sparsity of the abundance matrix. In this work, each spectral library element is appended to the data, and the abundance of the appended data is calculated assuming the whole spectral library as endmember by SUnSAL algorithm\cite{bioucas2009variable}. If the spectral library element is an image endmember, sparsity level of the resulting abundance matrix is lower as compared to sparsity of the original abundance matrix. On the other hand, if the spectral library endmember is not present in the image, sparsity level and numerical rank of the abundance matrix obtained after addition of the library element do not undergo significant change. However, estimation of sparsity level and numerical rank of the abundance matrix is a non-trivial task. The singular value plot as well as the numerical rank of the abundance matrix changes significantly depending on the presence of library element in the image.  This paper uses nuclear norm to identify the presence of library element, because, nuclear norm of a matrix is a surrogate of numerical rank. It is a convex function and it is the best convex approximation of rank of a matrix. The difference in nuclear norm of the actual spectral library abundance matrix and the abundance matrix obtained after addition of a spectral library element to the image gives a measure of change in numerical rank. As a result, nuclear norm of the appended abundance matrix undergo significant change for signal components and differ by an insignificant amount for non-signal components.
The proposed unmixing work flow is comprised of three stages-
\begin{itemize}
\item{Noise removal by multi linear regression}
\item{Abundance calculation by SUnSAL\cite{bioucas2009variable}}
\item{Dictionary pruning by nuclear norm maximization}
\end{itemize}

\subsection{Noise Removal by Multi Linear Regression}
Abundance calculation and corresponding nuclear norm maximization stages are affected by the presence of noise and outliers. As a consequence, efficient noise estimation and removal is desired before performing these stages. 

Multi linear regression based noise removal is extremely useful in denoising hyperspectral image. Spectral reflectance of consecutive spectral bands are highly correlated and hence, a particular spectral band can be modeled as a linear combination of the other spectral bands. 

The reflectance value of all pixels in the $i$-th band can be represented by- 
 \begin{equation}
 X_{:,i}=\beta_iY_{\sigma_i}+\xi_{:,i}
 \end{equation}
 where,$ X_{:,i}$ represents the reflectance profile of the $i$-th band
  \newline
  $\beta_i$ is the regression coefficient 
\newline  
  $Y_{\sigma_i}=[x_1,x_2,.,x_{i-1},x_{i+1},..x_L]$ is the reflectance of all bands except the i-th band.
  \newline
  and $\xi_{:,i}$ represents noise in the i-th band
 Now, the regression coefficient is calculated by-
 \begin{equation}
\tilde{\beta_i}=X\left ( Y_{\sigma_{i}}^{T} Y_{\sigma_{i}}\right )^{-1} Y_{\sigma_{i}}
 \end{equation}
 The noise in the i-th band can be estimated as-
\begin{equation}
\xi_{:,i}=X_{:,i}-\beta_iY_{\sigma_i}
\end{equation}
The noisefree image at the i-th band can be obtained by-
\begin{equation}
 \tilde{x_{:,i}}=x_{:,i}-\sigma_i
\end{equation}
The noise free data is taken for performing the algorithms.
\subsection{Abundance Calculation by SUnSAL}
Each pixel in a hyperspectral image consists of very few image endmembers. As a consequence, the abundance of the spectral library is highly sparse.  Traditional Moore-Penrose Pseudo inverse process based calculation of pseudo-inverse do not result in sparse solution. Constrained Spectral Unmixing by variable Splitting and Augmented Lagrangian (SUnSAL) \cite{bioucas2009variable} solves this problem by formulating it in optimization framework. It minimizes the reconstruction error term as well as minimizing sparsity of the reconstructed abundance matrix. It employs $L_1$ norm as the sparsity terms as $L_1$ norm is a measure of sparsity of the obtained abundance matrix. The optimization problem is solved by alternating direction method of multipliers (ADMM), because, it has good convergence property.

SUnSAL algorithm\cite{bioucas2009variable} minimizes the optimization function using adaptive multipliers-
\begin{equation}
\underset{M}{min} \left \| X-MD \right \|_{F}^{2}+ \lambda \left \| M \right \|_{1}
\end{equation}
Where, $\lambda>0$ is the lagrangian multiplier. The first term minimizes the reconstruction error, while the second term maximizes sparsity of the obtained abundance matrix.
\subsection{Dictionary Pruning by Nuclear Norm Maximization}
Let, $M$ be the abundance matrix obtained by SUnSAL algorithm \cite{tang2015sparse} considering the whole spectral library as endmember. The abundance matrix $M$ can be decomposed by singular value decomposition as-
\begin{equation}
M=U\Sigma V^{T}
\end{equation}
Where, $\sigma_{i}$ be the diagonal entries of $\Sigma\ $, and are the singular values of $M$.
Nuclear norm of the original spectral library abundance matrix is given by-
\begin{equation}
\left \| M \right \|_{*}=\sum_{i=1}^{min(N,L)}\sigma _{i}\left ( M \right )
\end{equation}
%Let, the singular value decomposition of the abundance matrix $\hat{M_i} $ obtained by appending $i$-th spectral library element to the data be given by-
%\begin{equation}
%\hat{M_i}=\hat{U}_i \hat{\Sigma}_i \hat{V}_i^{T}
%\end{equation}
%Nuclear norm of $\hat{M_i}$ is calculated by- 
%\begin{equation}
%\left \| \hat{M_i} \right \|_{*}=\sum_{i=1}^{min(N,L)}\sigma _{i}\left ( \hat{M_i} \right )
%\end{equation}
%The difference in the nuclear norm is-
%\begin{equation}
%\delta_i=\left \| M \right \|_{*}-\left \| \hat{M_i} \right \|_{*}
%\end{equation}

The spectral library elements having higher $\delta_i$ are more likely to be present in the data. The abundance matrix is calculated using SUnSAL method \cite{bioucas2009variable} that calculates the abundance of the spectral library by sparse approximation method. The advantage of using nuclear norm as the surrogate of numerical rank is that the nuclear norm is a convex function and produces result that has minimum cardinality.

Algorithm: Dictionary Pruning by Nuclear Norm Maximization
\begin{algorithm}
 \SetAlgoLined
 \KwIn{\bf{HSI Data $X$, Spectral Library \bf{$D$}, Number of endmembers $P$}}
 \KwOut{\bf{Pruned Spectral Library \bf{$\hat{D}$}}}
 \begin{itemize}
 \item{Calculate the abundance of the original image considering the whole spectral library as endmember using SUnSAL. Estimated abundance is $M$} 
 \item{Calculate the nuclear norm of the estimated abundance $M$. The nuclear norm is $\left \| M \right \|_{*}=\sum_{i=1}^{min(N,L)}\sigma _{i}\left ( M \right )$}
 \item{For each spectral library elements}\\
 \For{$i\leftarrow 1$ \KwTo $K$}{
 \begin{itemize}
 \item{Merge the $i$-th spectral library element with the data $X$. The merged  data $Y_{i}=\left [X;d_{i}\right]$}\\
 \item{Find the sparse abundance matrix of the appended data $Y_i$ using SUnSAL. The obtained abundance matrix $\hat{M_i}$}
 \item{Calculate the nuclear norm of $\hat{M_i}$. $\left \|\hat{M_i}\right \|_{*}=\sum_{i=1}^{min(N,L)}\sigma _{i}\left ( \hat{M_i} \right )$}
 \item{Calculate the difference between nuclear norm of abundance of the initial image and abundance of the appended image.\\
 $\delta_i=\left \| M \right \|_{*}-\left \| \hat{M_i} \right \|_{*}$}
 \end{itemize}
 }
 \item{The $P$ elements corresponding to the maximum $\delta_i$ are the actual spectral library endmembers present in the image}\\
 \item{Index of the pruned spectral library $\phi$}\\
 \item{Pruned spectral library $\hat{D}=d_\phi$}\\  
 \end{itemize}
\end{algorithm}

\begin{figure*}[!t]
 \center
  \includegraphics[width=9cm]{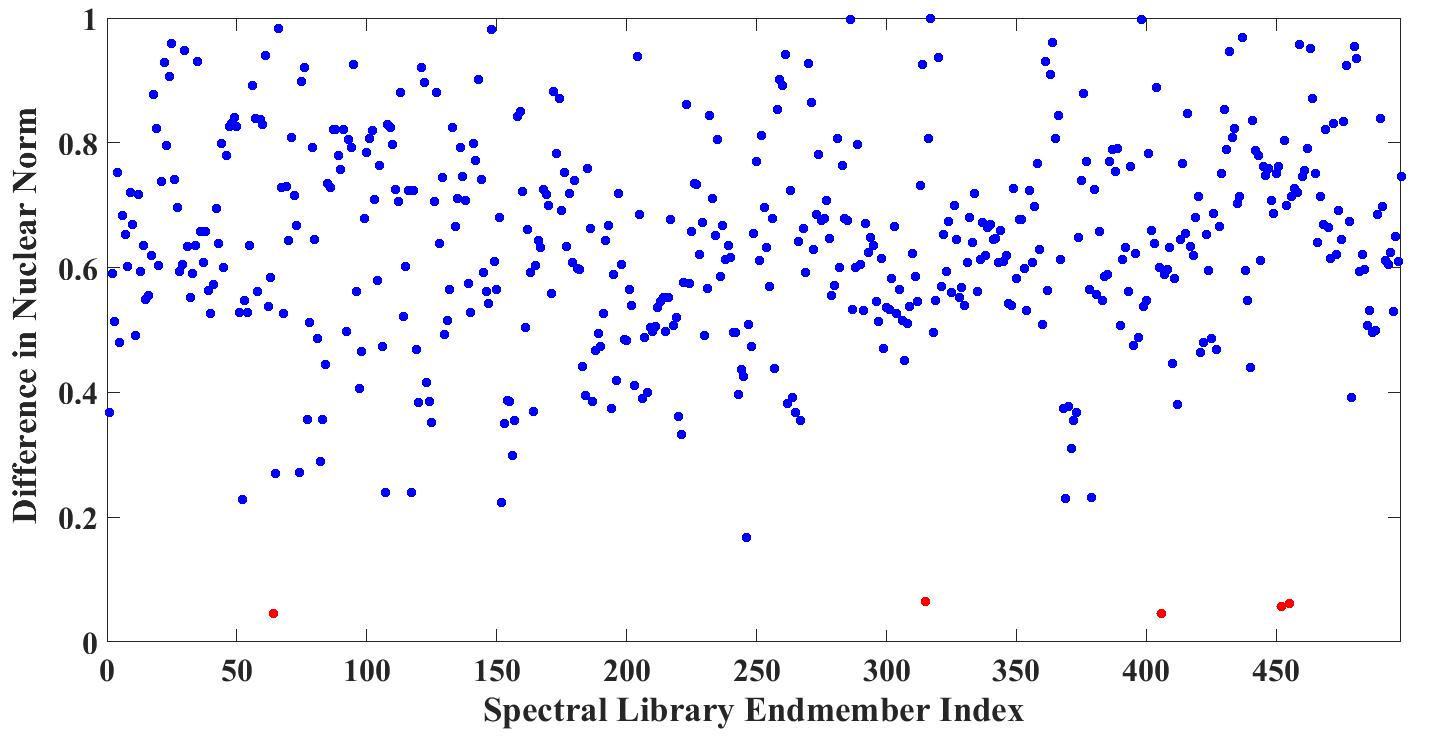}
  \caption{Spectral Library Endmember Index Vs Difference in Nuclear Norm of Abundance Matrix. Red dots indicate actual spectral library endmembers}
  \label{AAA}
\end{figure*}
\begin{figure*}[!t]
 \center
  \includegraphics[width=4cm]{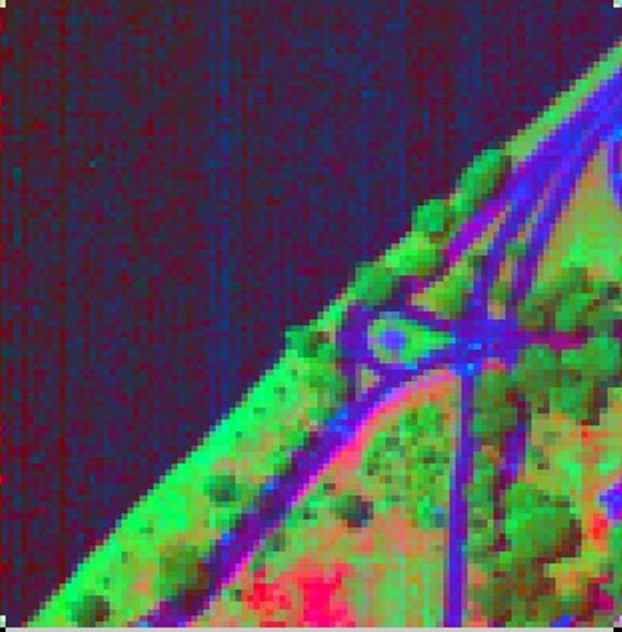}
  \caption{Washington DC Mall Image Displayed as RGB Combination of Band 1,8 and 16. }
  \label{AAA}
\end{figure*}
\begin{figure*}
 %\begin{multicols}{2}
\begin{centering}
 \begin{subfigure}[1]{0.5\textwidth}
    \includegraphics[width=8cm]{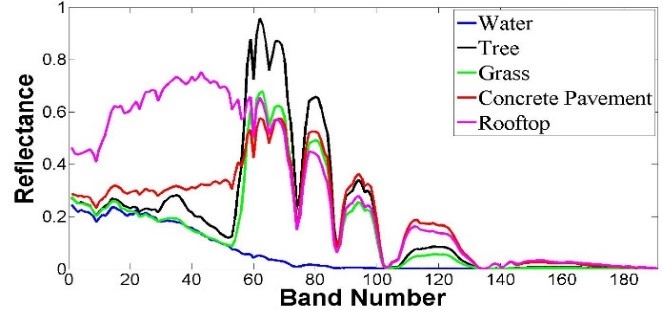}
    \caption{Actual Endmembers}
    \label{fig:gull}
 \end{subfigure}%
 \begin{subfigure}[2]{0.5\textwidth}
    \includegraphics[width=8cm]{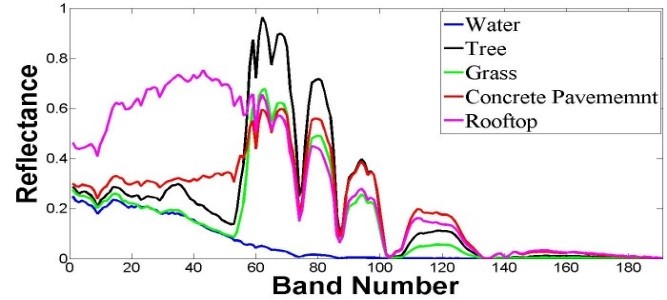}
    \caption{Endmembers Estimated by Nuclear Norm Difference based Dictionary Pruning}
    \label{fig:gull2}
 \end{subfigure}%
\end{centering}
 \caption{Spectral Reflectance of DC Mall Endmembers}
 %\end{multicols}
\end{figure*}
\section{Results}
The proposed unmixing methods were applied on a large number of synthetic as well as real images. Different situations like- noise level, pixel purity level, mutual coherence of the spectral library and number of endmembers were varied in these synthetic image experiments. 

\subsection{Performance Measures}
Performance of the unmixing methods have been measured by two parameters- average spectral distance (ASAD), signal to reconstruction error (SRE). Two other performance measures detection probability and probability of success are used to evaluate the performance of the dictionary pruning methods. 
\begin{itemize}
\item{\bf{Average Spectral Angle Distance (ASAD)}}\\
Spectral angle distance between actual and estimated endmembers is calculated by-
\begin{equation}
sad\left (s_{i},\hat{s_{j}} \right )=cos^{-1}\left (\frac{\left \langle s_{i},\hat{s_{j}} \right \rangle}{\left \|s_{i} \right \|_{2}{\left \| \hat{s_{j}} \right \|_{2}} } \right )
\end{equation}

\item{\bf{Signal to Reconstruction Error (SRE)}}\\
Signal to reconstruction error (SRE) denotes the relative power of reconstructed data with respect to the actual data. 
\begin{equation}
SRE=10\ log_{10}\left ( \frac{\left \| X \right \|_{2}}{\left \| X-\hat{X}\right \|_{2}}\right)
\end{equation}
where, $\hat{X}$ is the hyperspectral data reconstructed by the unmixing or dictionary pruning algorithm.

\item{\bf{Probability of Detection}}\\
Probability of detection defines the number of spectral library endmembers accurately selected. Probability of detection is defined as-
\begin{equation}
Pr\left(\Lambda\subset\hat{\Lambda}\right)
\end{equation}
where, $\Lambda$ is the indices corresponding the actual spectral library elements present in the image and $\hat{\Lambda}$ is the indices corresponding to the estimated spectral library elements of the pruned library. 
\end{itemize}
\subsection{Synthetic Image Experiments}
In this section synthetic data is generated by chosing randomly selected elements from USGS library. The abundance matrix was created such a way that it satisfies the convex constraints related to abundance (ASC and ANC). The dictionary pruning algorithms must not be influenced by- number of image pixels, number of image endmemeber and the maximum abundance of a particular endmember in a pixel. However, presence of noise degrades unmixing performance. 
In the first experiment, a synthetic hyperspectral data is generated from five spectral library endmembers. The existing dictionary pruning as well as unmixng algorithms are applied on this data. It can be seen from Fig. 1. that the library endmember indices corresponding to the minimum nuclear norm difference. 

\subsubsection{ Robustness to Number of Pixel and Abundance of Endmember}
In this experiment, the data is generated from five endmembers, but, the number of image pixels and the maximum abundance of an endmember in a pixel is varied. Here, the number of pixels is varied from 1000 and 500, whereas, the maximum abundance is changed from 1 to 0.6. It is evident from Table I that MUSIC-CSR\cite{iordache2014music} gives the best performance in terms of SRE, our proposed nuclear norm difference result in more or less satisfactory result. However, nuclear norm method has edge over MUSIC-CSR and other methods in terms selecting optimum endmemeber set. Table III indicates that nuclear norm difference method results in higher detection probability.

\subsubsection{Robustness to Number of Endmembers and Noise Level}
In this experiment, both the number of image endmebers and the noise level are varied. The number of image endmembers is varied from 5 to 10. The result shown in Table II shows that the proposed nuclear norm difference maximization gives good SRE, however, MUSIC-CSR\cite{iordache2014music} gives the best performance. However, it is evident from Table IV that nuclear norm difference scheme detects the optimum endmembers, hence, has very high detection probability.

\subsection{Real Image Experiment}
Our proposed unmixing algorithm is validated on HYDICE Washington DC Mall image, which is extensively used for analyzing performance of unmixing. This image is taken from HyDICE sensor and it consists of a total 210 spectral bands at 10 nm spectral resolution in the electromagnetic wavelength range 0.4-2.4 µm region. However, due to mismatch between USGS spectral library and the image, both the image and the library has been calibrated. The calibrated version contains 182 spectral bands. Previous ground truth study reports that the image contains five pre-dominant endmembers- tree, water, grass, concrete pavement and rooftop. Algorithms for estimating the number of endmembers such as- HFC-VD\cite{chang2004estimation}, HYSIME\cite{bioucas2008hyperspectral} also show similar result. The actual and estimated spectral library endmembers are displayed in Figure 3.a) and b) respectively.
Table V shows that though the SRE value obtained by nuclear norm difference based method is comparable to other methods, but, it has higher detection probability.

\section{Conclusion and Future Scope}
This paper proposes a new dictionary pruning algorithm, which is based on nuclear norm difference maximization. The experimental results show that the proposed algorithm gives good performance in different conditions, such as- in the presence of low-degree mixed pixels, lower number of endmembers and in the presence of noise. Nuclear norm based approach overcomes the requirement of estimating rank of the data because, it is a convex surrogate of rank of a matrix.

The proposed workflow is applicable to images which can be approximated as linear mixture model. This work can be extended for non-linear dictionary pruning.

\bibliography{Biblio}
\bibliographystyle{IEEEtran}
%\bibliography{A}
\end{document}